\documentclass{emulateapj}

\slugcomment{Accepted to the Astrophysical Journal} 

\shorttitle{The \lowercase{$r$}-Process and Termination Shock}

\shortauthors{Kuroda}

\begin{document}

\title{The \lowercase{$r$}-Process in Supersonic Neutrino-Driven
Winds: The Roll of Wind Termination Shock}

\author{Takami Kuroda, Shinya Wanajo, and Ken'ichi Nomoto}

\affil{Department of Astronomy, School of Science,
   University of Tokyo, Bunkyo-ku, Tokyo, 113-0033, Japan;
   kuroda@astron.s.u-tokyo.ac.jp, wanajo@astron.s.u-tokyo.ac.jp,
   nomoto@astron.s.u-tokyo.ac.jp}

\begin{abstract}
Recent hydrodynamic studies of core-collapse supernovae imply that the
neutrino-heated ejecta from a nascent neutron star develops to
supersonic outflows. These supersonic winds are influenced by the
reverse shock from the preceding supernova ejecta, forming the wind
termination shock. We investigate the effects of the termination shock
in neutrino-driven winds and its roll on the $r$-process. Supersonic
outflows are calculated with a semi-analytic neutrino-driven wind
model. Subsequent termination-shocked, subsonic outflows are obtained
by applying the Rankine-Hugoniot relations. We find a couple of
effects that can be relevant for the $r$-process. First is the sudden
slowdown of the temperature decrease by the wind termination. Second
is the entropy jump by termination-shock heating, up to several $100\,
N_\mathrm{A} k$. Nucleosynthesis calculations in the obtained winds
are performed to examine these effects on the $r$-process. We find
that 1) the slowdown of the temperature decrease plays a decisive roll
to determine the $r$-process abundance curves. This is due to the
strong dependences of the nucleosynthetic path on the temperature
during the $r$-process freezeout phase. Our results suggest that only
the termination-shocked winds with relatively small shock radii ($\sim
500$~km) are relevant for the bulk of the solar $r$-process abundances
($A \approx 100-180$). The heaviest part in the solar $r$-process
curve ($A \approx 180-200$), however, can be reproduced both in
shocked and unshocked winds. These results may help to constrain the
mass range of supernova progenitors relevant for the $r$-process. We
find, on the other hand, 2) negligible roles of the entropy jump on
the $r$-process. This is a consequence that the sizable entropy
increase takes place only at a large shock radius ($\ga 10,000$~km)
where the $r$-process has already ceased.
\end{abstract}


\section{Introduction}
Whether the rapid-neutron capture ($r$-process) can take place under
the circumstance of core-collapse supernovae has been an unresolved
problem for a long time \citep[see][for a recent
review]{Cowa04,Wana06c, Arno07}. The requisite physical conditions
include neutron-rich environment, high entropy, and short expansion
timescale to obtain the high enough neutron-to-seed abundance ratio
($\sim 100$) for the production of heavy $r$-process nuclei. The
neutrino-heated supernova ejecta (neutrino-driven winds) from a
proto-neutron star has been expected as a suitable astrophysical site
to fulfil such physical conditions \citep[in particular high entropy
and short expansion timescale,][]{Woos92, Meye92, Woos94, Taka94,
Qian96, Card97, Otsu00, Sumi00, Wana01, Thom01}.

Previous studies confirmed, however, that the spherically expanding
winds from a typical proto-neutron star (e.g., $1.4\, M_\odot$ with a
10~km radius) cannot attain the requisite physical conditions for the
$r$-process. Some mechanisms to cure this problem have been proposed,
e.g., general relativistic effects \citep{Card97, Otsu00, Wana01,
Thom01}, magnetic field \citep{Thom03, Suzu05}, anisotropic neutrino
radiation \citep{Wana06b}, and acoustic heating \citep{Burr07},
however, no consensus has been achieved. Recently, \citet{Arco07} have
suggested that the reverse shock propagating from the preceding
supernova ejecta might have important effects on the $r$-process
nucleosynthesis.  Two-dimensional hydrodynamic studies of
core-collapse supernovae have shown that the outflow driven by
neutrino heating develops to be supersonic, which eventually
decelerate by the reverse shock from the outer layers
\citep[e.g.,][]{Jank95, Jank96, Burr95, Bura06}.  \citet{Arco07} have
explored the effects of the reverse shock on the properties of
neutrino-driven winds by one-dimensional, long-time hydrodynamic
simulations of core-collapse supernovae. They have found that, in
their all models ($10-25\, M_\odot$ progenitors), the outflows become
supersonic and form the ``termination shock'' when colliding with the
slower preceding supernova ejecta. This condition continues until the
end of their computations (10 seconds after core bounce) in their all
``standard'' models (with reasonable parameter choices). Their results
show that the entropy abruptly increases and the temperature continues
to decrease very slowly in the termination-shocked
winds. \citet{Arco07} have suggested that the strongly time-dependent
behaviours of the shocked conditions would play a non-negligible role
to the $r$-process nucleosynthesis.

Systematic nucleosynthesis calculations with the termination-shocked,
supersonic neutrino-driven winds, however, have been absent, despite a
number of works with the subsonic (``breeze'') outflows
\citep[e.g.,][]{Wana01}. \citet{Thom01} briefly discussed possible
effects of the shock-decelerated winds on the $r$-process
abundances. They calculated, however, only one specific case with an
arbitrary chosen shock radii, in which the entropy increase was only
moderate ($\sim 10\, N_\mathrm{A} k$). \citet{Wana02} and
\citet{Wana07} explored the effects of the terminated supersonic winds
on the $r$-process by introducing a constant ``freezeout temperature''
$T_\mathrm{f}$. They showed the impact of $T_\mathrm{f}$ on the
$r$-process abundance curves, in particular, near the third
$r$-process peak ($A \approx 195$) and beyond. However, $T_\mathrm{f}$
was taken to be an arbitrary chosen free parameter. Their approach
only roughly mimicked the wind-termination effects, and the resulting
entropy increase was not taken into account.

In this study, we investigate the effects of the wind termination
shock on the $r$-process nucleosynthesis in some detail. For this
purpose, the thermodynamic histories of supersonic outflows are
calculated with a semi-analytic, spherically symmetric, general
relativistic neutrino-driven wind model \citep{Wana01, Wana02}. The
termination-shocked outflows are then obtained by applying the
Rankine-Hugoniot relations at arbitrary chosen shock radii (\S~2). In
\S~3, the behaviors of the shock-heated winds are presented along with
their dependencies on the neutron star masses, the neutrino
luminosities, and the shock radii. By applying the obtained wind
trajectories, the $r$-process calculations are performed with the
extensive nuclear reaction network (\S~4). We discuss effects of the
termination-shock deceleration and heating on the $r$-process
nucleoynthesis. Our conclusions of this study and discussion are
presented in \S~5.

\section{Neutrino-Driven Wind Model}

\subsection{Wind Equations}

After the gravitational collapse of a massive star ($> 8\, M_\odot$),
a great amount of neutrinos are emitted from the proto-neutron star.
Beyond the neutrino-sphere, where the neutrino opacity becomes $\la1$,
heating of matter by neutrinos dominates over the cooling
process. Matter in the vicinity of the neutrino sphere is then blown
off as ``neutrino-driven winds''. The physical profiles of each wind
can be obtained by solving the wind equations \citep{Dunc86, Qian96,
Card97, Otsu00, Wana01, Thom01}.

Steady state spherically symmetric wind equations including general
relativistic effects are described by the following mass, momentum and
energy conservations,
\begin{equation}
\dot{M}=4\pi r^2 \rho u,
\end{equation}
\begin{equation}
u\frac{du}{dr}=-\frac{1+(u/c)^2-2GM_\mathrm{NS}/rc^2}{\rho(1+\varepsilon/c^2)+P/c^2}\frac{dP}{dr}-\frac{GM_\mathrm{NS}}{r^2},
\end{equation}
\begin{equation}
\dot{q}=u\left(\frac{d\varepsilon}{dr}-\frac{P}{\rho^2}\frac{d\rho}{dr}\right),
\end{equation}
where $r, \rho, P, \varepsilon$, $\dot{M}$, and $M_\mathrm{NS}$ denote
the distance from the center, matter density, pressure, specific
internal energy, mass ejection rate from the surface, and
gravitational mass of the proto-neutron star, respectively. The radial
velocity $u$ is related to the proper velocity of the matter $v$ as
measured by a local, stationary observer by $v = u/\sqrt{1 + (u/c)^2 -
2GM_\mathrm{NS}/rc^2}$. The specific heating rate $\dot{q}$ includes
both heating and cooling by neutrino interactions described in
\citet{Otsu00}, in which general relativistic effects are explicitly
taken into account. In this study, the neutrino luminosities $L_\nu$
of all flavors are assumed to be equal, and the rms average neutrino
energies are taken to be 10, 20, and 30~MeV for electron,
anti-electron, and the other flavors of neutrinos, respectively.

To close the above equations~(1)-(3), equations of state have to be
added. We assume that the wind matter is composed of relativistic
electrons, positrons and photons, and also non-relativistic free
nucleons. The equations of state can be then described as
\begin{equation}
P=\frac{11\pi^2}{180}\frac{k^4}{c^3\hbar^3}\frac{T^4}{\rho}+\frac{\rho kT}{m_N},
\end{equation}
\begin{equation}
\varepsilon=\frac{11\pi^2}{60}\frac{k^4}{c^3\hbar^3}T^4+\frac{3}{2}\frac{kT}{m_N},
\end{equation}
where $m_N$ is the nucleon rest mass and $T$ is the matter
temperature. Effects of arbitrary relativity and degeneracy of
electrons are ignored, which do not significantly modify the
thermodynamic histories of winds \citep[e.g., up to 10\% in asymptotic
entropy,][]{Wana01}. Solving equations~(1)-(5) with boundary
conditions, we can obtain the thermodynamic trajectories $u(r)$,
$T(r)$, and $\rho(r)$.

The radius of the proto-neutron star $R_{\rm NS}$ is taken to be
10~km, which is assumed to be equal to the neutrino sphere.  The
boundary conditions at $R_\mathrm{NS}$ are set to be $\rho
(R_\mathrm{NS}) = 10^{10}$~g~cm$^{-3}$ and $T(R_\mathrm{NS})$ that
equalizes the neutrino heating rate with cooling rate. The velocity
$u(R_\mathrm{NS})$ (or equivalently $\dot{M}$ in eq.~[1]) is
determined by an iterative relaxation method with equations~(1)-(5)
and the above boundary conditions to obtain the transonic (i.e.,
becoming supersonic through the sonic point) wind solutions.

\subsection{Termination-Shock Conditions}

In each transonic wind, we assume that a termination shock appears at an
arbitrarily radius $R_\mathrm{s}$ beyond the sonic radius
$R_\mathrm{c}$, at which the Rankine-Hugoniot shock jump conditions
are applied.
The Rankine-Hugoniot equations for mass, momentum, and energy
conservations are
\begin{equation}
\rho_1 u_1=\rho_0 u_0,
\end{equation}
\begin{equation}
P_1+\rho_{1}u_{1}^2=P_0+\rho_{0}u_{0}^2,
\end{equation}
\begin{equation}
\frac{1}{2}{u_1}^2 + \varepsilon_1 + \frac{P_1}{\rho_1}
= \frac{1}{2}{u_0}^2 + \varepsilon_0 + \frac{P_0}{\rho_0},
\end{equation}
where the subscripts 1/0 denote quantities in shocked/unshocked
outflows just above/behind $R_\mathrm{s}$. We neglect the relativistic
effects (i.e., $u = v$) because of the distant shock positions
($R_\mathrm{s} > R_\mathrm{c}$) from the surface,
where $u$ $\la 0.1 c$ (see Figs.~1 and 2). In
equations~(6)-(8), the velocity of the shock position is assumed to be
negligible compared to the wind velocity ($\sim 10^9\, \rm {cm \
s^{-1}}$). The termination-shocked subsonic solutions are then
calculated by solving equations~(1)-(5) with the boundary conditions
just above $R_\mathrm{s}$ obtained from equations~(6)-(8) (with
eqs.~[4] and [5]).


\section{Properties of Termination-Shocked Winds}

The solutions of termination-shocked winds $u$ (\textit{top left}),
$T$ (\textit{top right}), and $\rho$ (\textit{bottom left}), together
with the entropy per baryon $s$ (\textit{bottom right}), as functions
of $r$ are displayed in Figure~1. The neutron star mass is taken to be
$M_{\rm{NS}}=1.4\, M_\odot$. The neutrino luminosities are assumed to
be $L_{\nu,51} = 10$ (\textit{blue lines}) and 1 (\textit{green
lines}), where $L_{\nu,51}\equiv L_\nu /10^{51}$~erg. The former and
latter can be regarded as representative of the early and late winds,
respectively \citep[$\sim 1-2$~s and $\ga 10$~s after core
bounce,][]{Woos94, Arco07}. In the hydrodynamic results by
\citet{Arco07}, the termination shock appears at $R_\mathrm{s} =
3000-7000$~km in their fiducial model (for the $15\, M_\odot$
progenitor star), but $R_\mathrm{s}$ is highly progenitor dependent as
well as time dependent, ranging from a few $100$~km to several
10000~km. We consider, therefore, various cases with $R_\mathrm{s} =
500$, 1000, 3000, 10000, and 30000~km, whenever $R_\mathrm{s} >
R_\mathrm{c}$ is satisfied. For comparison, the cases with
$R_\mathrm{s} = \infty$ and $R_\mathrm{s} = R_\mathrm{c}$ are also
displayed in Figure~1 (sonic points are denoted by
\textit{stars}). The former corresponds to the unshocked transonic
solution. In the latter case, the shock jump disappears and thus the
wind is smoothly connected to the subsonic breeze solution through the
sonic point, with the maximum $u(R_\mathrm{NS})$ (i.e., maximum
$\dot{M}$).

As seen in Figure~1, supersonic winds abruptly decelerate at
$R_\mathrm{s}$ (\textit{top left}) to become subsonic breeze flows. As
a consequence, the temperature (\textit{top right}) and density
(\textit{bottm left}) decrease rather slowly after the shock jumps,
and the entropy increases by the termination-shock heating. The shock
jump becomes large as the shock position moves away from the sonic
point. In particular, the extreme entropy jump is seen in the distant
$R_\mathrm{s}$ ($\ga 10000$~km) wind, up to $\sim 400\, N_\mathrm{A}
k$ in the highest case. At a fixed $R_\mathrm{s}$, the early wind
($L_{\nu, 51} = 10$) results in a larger entropy jump, which has the
larger kinetic energy to be converted to thermal energy (Fig.~1;
\textit{top left}). The termination-shocked entropies in the early
winds are appreciably higher, despite their significantly lower
unshocked entropies than those in the late winds.

The hydrodynamic study by \citet{Arco07} shows variations of
$M_\mathrm{NS}$ with time as well as depending on the progenitor
masses, ranging from $1.1\, M_\odot$ to $2.0\, M_\odot$. In Figure~2,
the wind solutions for $M_\mathrm{NS} = 2.0\, M_\odot$ (\textit{red
lines}) as an extreme case are compared to those for $M_{\rm NS}=
1.4\, M_\odot$ (\textit{green lines}), in which the neutrino
luminosity is fixed to $L_{\nu ,51}=10$ (otherwise the same as
Fig.~1).

As can be seen, the unshocked asymptotic entropies for $M_{\rm
NS}=2.0M_\odot$ are about a factor of two higher than those for
$M_\mathrm{NS} = 1.4\, M_\odot$ (Fig.~2, \textit{bottom left}) owing
to general relativistic effects \citep{Card97, Otsu00, Wana01,
Thom01}. In addition, we find larger entropy jumps for $M_\mathrm{NS}
= 2.0\, M_\odot$ because of the somewhat faster radial velocities
(Fig.~2, \textit{top left}). As a consequence, the termination-shocked
entropies for $M_\mathrm{NS} = 2.0\, M_\odot$ are appreciably higher
than those for $M_\mathrm{NS} = 1.4\, M_\odot$.

All the above aspects are qualitatively similar to those found in the
hydrodynamic results by \citet{Arco07}.
For example, in their fiducial model (with $M_\mathrm{NS} = 1.2-1.3\,
M_\odot$ resulting from the $15\, M_\odot$ progenitor), the unshocked
and shocked entropies at $2.0$~s after core bounce ($L_{\nu, 51} \sim
10$ and $R_\mathrm{s} \sim 4000$~km) are $\sim 70\, N_\mathrm{A} k$
and $\sim 180\, N_\mathrm{A} k$, respectively. This is in reasonable
agreement with our result in the wind with $M_\mathrm{NS} = 1.4\,
M_\odot$, $L_{\nu, 51} = 10$, and $R_\mathrm{s} = 5000$~km (not
displayed in Figs.~1 and 2 but shown in Fig.~3), in which the unshocked and
shocked entropies are $76\, N_\mathrm{A} k$ and $166\, N_\mathrm{A}
k$, respectively. Agreements to a similar extent can be found in other
relevant cases.
Our semi-analytic approach is therefore quite useful to obtain the
thermodynamic histories of termination-shocked winds, which are
consistent with detailed hydrodynamic results at qualitative levels.

We note that the termination-shocked outflow merges with the preceding
supernova ejecta and it cannot be regarded as a wind any more. In
fact, the hydrodynamic calculations show that the shocked outflows
further decelerate to be nearly constant velocities \citep[e.g.,
Fig.~5 in][]{Arco07} when accumulated in a dense shell between the
forward and reverse shocks. Nevertheless, we apply the wind solutions
for the termination-shocked outflows as described in \S~2. As a
consequence, the outflow velocity rapidly drops (Figs.~1 and 2),
resulting in the slower decreases of temperature and density with time
(Fig.~5) than those in the hydrodynamic calculations \citep[Fig.~6
in][]{Arco07}. It should be noted, however, that the $r$-process
nucleosynthesis ceases only within $\sim 0.2-0.3$~s after (or, in some
cases, even before) the wind termination in the current
nucleosynthesis calculations \citep[\S~4, see also][]{Wana07}. Our
treatment may be thus reasonable for the current purpose (i.e., for
nucleosynthesis), since the time variations of temperature and density
during such a short timescale would be negligible in the
shock-decelerated outflows.

Figure~3 shows relations between the entropies $s_{\rm s}$ and the
temperatures $T_{\rm 9,s}$ just above the shock radii for various
$R_\mathrm{s}$ cases. The neutron star masses are taken to be $M_{\rm
NS} = 1.4\, M_\odot$ (\textit{crosses}) and $2.0\, M_\odot$
(\textit{asterisks}). The results with various $R_{\rm s}$ (same as in
Figs.~1 and 2) are connected with solid and dot-dashed lines for
$L_{\nu,51}=10$ and 1, respectively. The cases of $R_\mathrm{s} =
R_\mathrm{c}$, 1000~km, and 10000~km are marked by circles,
squares, and triangles, and the numbers indicated at the circles are
the sonic radii $R_\mathrm{c}$. Note that the entropy just behind the
shock for each case is the same as $s_\mathrm{s}$ in the $R_\mathrm{s}
= R_\mathrm{c}$ (subsonic) wind (marked by \textit{circles}).
As seen in Figure~3, the entropy jump occurs only when the temperature
decreases below $T_9 \approx 2.5$ (at which the $r$-process
begins). We find a trend that $T_\mathrm{9,s}$ is systematically
higher in earlier (i.e., higher $L_\nu$) winds, while $M_\mathrm{NS}$
is less sensitive to $T_\mathrm{9,s}$. As an extreme case, we also
display the results with $M_{\rm NS}=1.4\, M_\odot$, $R_{\rm
{NS}}=10$~km, and $L_{\nu,51}=70$ (\textit{dashed line}). In these
cases, the entropy jump occurs at a smaller $R_\mathrm{s}$ and thus at
significantly higher temperature, but the resulting entropy is only
moderate ($\la 100\, N_\mathrm{A} k$) for $T_\mathrm{9, s} > 2.5$. In
addition, the neutrino sphere at earlier times would be significantly
larger than 10~km and thus the entropy would be lower than the current
case.

Moreover, matter in the earlier winds with $L_{\nu, 51} > 10$ would
be proton-rich \citep{Bura06, Arco07}, in which no
$r$-processing is expected.

We conclude, therefore, the entropy jump by termination-shock heating
does not help to enhance the neutron-to-seed ratio prior to the
$r$-process ($T_9 > 2.5$). The high entropy should be attained much
earlier than the current situation to influence the neutron-to-seed
ratio \citep{Qian96, Suzu05, Wana06b, Burr07}. The termination shock
can, however, play a decisive role to determin the $r$-process curve
as described in \S~4.

\section{Nucleosynthesis}

\subsection{Nuclear Reaction Network}

Adopting the thermodynamic trajectories discussed in \S~2 for the
physical conditions, the nucleosynthetic yields are obtained by
solving an extensive nuclear reaction network code. The network
consists of 6300 species between the proton and neutron drip lines
predicted by a recent fully microscopic mass formula
\citep[HFB-9,][]{Gori05}, all the way from single neutrons and protons
up to the $Z = 110$ isotopes. All relevant reactions, i.e. $(n,
\gamma)$, $(p,\gamma)$, $(\alpha, \gamma)$, $(p, n)$, $(\alpha, n)$,
$(\alpha, p)$, and their inverse are included. The experimental data
whenever available and the theoretical predictions for light nuclei
($Z < 10$) are taken from the
REACLIB\footnote{http://nucastro.org/reaclib.html. We used the older
version on the web, which has been recently updated with new
experimental rates.}  compilation. The three-body reaction $\alpha
(\alpha n, \gamma)^9$Be, which is of special importance as the
bottleneck reaction to heavier nuclei, is taken from the experimental
data of \citet{Utsu01}. All other reaction rates are taken from the
Hauser-Feshbach rates of
BRUSLIB\footnote{http://www.astro.ulb.ac.be/Html/bruslib.html.}
\citep{Aika05} making use of experimental masses \citep{Audi03}
whenever available or the HFB-9 mass predictions \citep{Gori05}
otherwise. The $\beta$-decay rates are taken from the gross theory
predictions \citep[GT2,][]{Tach90}, obtained with the HFB-9
$Q_{\beta}$ predictions (T. Tachibana 2005, private
communication). Neutrino-induced reactions as well as the nuclear
fission are not considered in this study.

Each calculation is initiated when the temperature decreases to $T_9 =
9$ (where $T_9 \equiv T/10^9\, \mathrm{K}$). At this high temperature,
the compositions in the nuclear statistical equilibrium are obtained
(mostly free nucleons and $\alpha$ particles) immediately after the
calculation starts. The initial compositions are thus given by $X_n =
1 - Y_{e}$ and $X_p = Y_{e}$, respectively, where $X_n$ and $X_p$ are
the mass fractions of neutrons and protons, and $Y_{e}$ is the initial
electron fraction (number of proton per nucleon) at $T_9 = 9$. In this
study, $Y_e$ is taken as a free parameter and varied from 0.20 to 0.50
with an interval of 0.01.

\subsection{Nucleosynthetic Abundances}

Figure~4 shows the resulting neutron-to-seed ratio $Y_n/Y_h$ at $T_9 =
2.5$ (approximately the beginning of $r$-processing) for each
($M_\mathrm{NS}$, $L_\nu$) set as a function of $Y_e$. Here, $Y_n$ and
\begin{equation}
Y_h \equiv \sum_{Z>2, A} Y(Z, A)
\end{equation}
are the abundances of free neutrons and the heavy nuclei ($Z > 2$),
respectively. The solid and dotted lines denote the results for
$L_{\nu, 51} = 10$ and 1, respectively. For $M_\mathrm{NS} = 1.4\,
M_\odot$, $Y_n/Y_h$ is only a few 10 at $Y_e \approx 0.35$ \citep[that
is the lowest $Y_e$ value observed in hydrodynamic
results,][]{Woos94}. In this condition, only the second $r$-process
peak ($A = 130$) can be formed \citep{Wana01}, and a significantly
lower $Y_e$ ($\sim 0.25$) is needed for the third peak ($A = 195$)
formation. On the other hand, for $M_\mathrm{NS} = 2.0\, M_\odot$,
$Y_n/Y_h \sim 100$ is obtained with a reasonable choice of $Y_e$ ($
\sim 0.4$), which is sufficient for the third peak ($A = 195$)
formation \citep{Wana01}. In the following, the results for
$M_\mathrm{NS} = 1.4\, M_\odot$ are presented, keeping in mind that
the values of $Y_e$ are significantly lower-shifted compared to the
hydrodynamic results \citep{Woos94, Arco07}. Note that qualitatively
similar nucleosynthetic results can be obtained for $M_\mathrm{NS} =
2.0\, M_\odot$ with higher-shifted $Y_e$ values.

As seen in Figure~4, the $Y_n/Y_h$ values in the early ($L_{\nu, 51} =
10$) and late ($L_{\nu, 51} = 1$) winds (in particular for
$M_\mathrm{NS} = 1.4\, M_\odot$) at the same $Y_e$ are not
significantly different, despite a large difference in entropy
(Fig.~1). This is due to the shorter expansion timescale in the early
wind, in which the entropy is lower. These two effects compensate with
each other and lead to similar nucleosynthetic results
\citep{Wana01}. We find, therefore, that the effect of $Y_e$ variation
(that is most difficult to constrain) on nucleosynthesis is much
larger than the $L_\nu$ history. In fact, only ten percent change in
$Y_e$ leads to a substantial difference in the $r$-process curve
(Figs.~6 and 7). Note that the high $Y_n/Y_h$ values at $Y_e =
0.48-0.50$ for $M_\mathrm{NS} = 2.0\, M_\odot$ with $L_{\nu, 51} = 10$
are due to its rather short expansion timescale, in which $Y_h$ is
significantly small \citep{Hoff97, Wana02}.

The time variations of $r$, $T_9$, $\rho$, and $s$ for $R_\mathrm{s} =
215$~km ($= R_\mathrm{c}$), 500, 1000, 3000, 10,000, and 30,000~km in
the early wind ($L_{\nu, 51} = 10$) with $M_\mathrm{NS} = 1.4\,
M_\odot$ are shown in Figure~5. The calculated abundance curves with
these trajectories are displayed in Figures~6 and 7. The assumed $Y_e$
values of $0.35 \pm 0.04$ (Fig.~6) and $0.25 \pm 0.04$ (Fig.~7) are
responsible for the second and third peak formations,
respectively. Such small variations in $Y_e$ with time would be
inevitable when realistic time evolution of neutrino-driven outflows
were considered \citep[e.g.,][]{Woos94, Bura06}. Roughly speaking,
therefore, the \textit{envelope} of these five curves in each panel
might be regarded as the mass-averaged abundance pattern expected in
more realistic, time-evolving winds \citep[see, e.g.,][]{Wana07}.

In Figure~6, the envelope for $R_\mathrm{s} = 500$~km best reproduces
the solar abundance pattern \citep[\textit{dots},][]{Kapp89} between
$A = 100$ and 180, including the second and rare-earth peak
abundances. The $R_\mathrm{s} = R_\mathrm{c} = 215$~km case forms the
troughs at the low ($A = 110-120$) and high ($A = 140-150$) sides of
the second peak. This is in fact occasionally attributed to the
problems in the predicted nuclear masses around $N = 82$
\citep{Wana04}. For $R_\mathrm{s} \ge 1000$~km, the envelopes
unacceptably broaden compared to the solar $r$-process curve. 

For the third peak abundances ($A = 180-200$ in Fig.~7), in contrast,
the envelopes except for $R_\mathrm{s} = R_\mathrm{c} = 215$~km are in
reasonable agreement with the solar $r$-process curve. In particular,
better fits are seen for $R_\mathrm{s} = 1000$~km and $R_\mathrm{s}
\ge 10,000$~km, while the intermediate $R_\mathrm{s}$ ($= 3000$~km)
case results in the lower-shifted third peak. Note that the abundance
curves for $R_\mathrm{s} \ge 3000$~km in Figure~6 and for
$R_\mathrm{s} \ge 10,000$~km in Figure~7 cannot be distinguished. All
these aspects can be understood as a consequence of the
termination-shock deceleration of supersonic winds, as discussed in
\S~4.3.

\subsection{Effect of Termination-Shock Deceleration}

The heavy nuclei synthesis ceases when the neutron-to-seed ratio
decreases to $Y_n/Y_h = 1$ \citep[hereafter,
``$n$-exhaustion'',][]{Wana04, Wana07}. Figure~8 (\textit{top left})
shows $T_9$ at $n$-exhaustion, $T_\mathrm{9,e}$, for each
$R_\mathrm{s}$ as a function of $Y_e$. In each line, a temperature
jump appears as $Y_e$ decreases (except for $R_\mathrm{s} =
R_\mathrm{c}$). As an example, the line for $R_\mathrm{s} = 500$~km
(\textit{red line}) overlaps with those for $R_\mathrm{s} > 500$~km at
$Y_e > 0.38$. This is a consequence that $n$-exhaustion occurs before
reaching $R_\mathrm{s}$ because of the low $Y_n/Y_h$ values
(Fig.~4). For $Y_e \le 0.38$, $n$-exhaustion takes place after the
termination-shock deceleration (with the temperature jump) owing to
the sufficiently high $Y_n/Y_h$ values. In these cases,
$T_\mathrm{9,e}$ is almost independent of $Y_e$ because of the rather
slow temperature decrease after the wind termination (Figs.~1 and
2). Note that, for $R_\mathrm{s} = 30,000$~km, $n$-exhaustion occurs
before the termination-shock deceleration in all $Y_e$ cases. As seen
in Figure~8 (\textit{top left}), the $T_\mathrm{9,e}$ curves for
$R_\mathrm{s} \ge 3000$~km are overlapped at $Y_e > 0.30$. Similarly,
the $R_\mathrm{s} \ge 10,000$~km cases result in the same
$T_\mathrm{9,e}$ curve for $Y_e > 0.21$. This is the reason why the
abundance curves in these cases cannot be distinguished (Figs.~6 and
7).

No heavier nuclei are synthesized after $n$-exhaustion, but local
rearrangement of the abundance distribution continues until the
neutron capture becomes slower than the competing $\beta$-decay. The
final $r$-process curve is thus fixed at $\tau_{n\gamma}/\tau_\beta =
1$ \citep[hereafter, ``freezeout'';][]{Wana04, Wana07}. Here,
\begin{equation}
\tau_{n\gamma} \equiv
\left[\frac{\rho Y_n}{Y_h} \sum_{Z>2, A} \lambda_{n\gamma} (Z, A)\, Y(Z, A) \right]^{-1}
\end{equation}
and
\begin{equation}
\tau_{\beta} \equiv
\left[\frac{1}{Y_h}\sum_{Z>2, A} \lambda_{\beta} (Z, A)\, Y(Z, A) \right]^{-1},
\end{equation}
are the abundance-averaged mean lifetimes of $(n, \gamma)$ reactions
and $\beta$-decays for $Z > 2$ nuclei, respectively, where
$\lambda_{n\gamma}(Z, A)$ and $\lambda_{\beta}(Z, A)$ are the rates of
the corresponding reactions. These represent the lifetimes of the
dominant species at a given time. The ``classical'' $r$-process is
characterized by the conditions $\tau_{n\gamma}/\tau_{\gamma n}
\approx 1$ and $\tau_\beta/\tau_n \gg 1$, where $\tau_{\gamma n}$ is
the $(\gamma, n)$ lifetimes defined similar to equation~(11).

The $r$-process path for each case can be specified in terms of the
neutron separation energies. In this study, 
the abundance-averaged $2n$ separation energy $S_{2n}$
divided by two, defined by
\begin{equation}
S_\mathrm{a} \equiv
\frac{1}{Y_h} \sum_{Z>2, A} \frac{S_{2n}(Z, A)}{2}\, Y(Z, A)
\end{equation}
\citep{Wana07}, is taken to represent the nucleosynthetic path at a
given time. That is, $S_\mathrm{a} = 0$ is the neutron-drip line,
while a higher $S_\mathrm{a}$ locates at closer to the
$\beta$-stability line. In Figure~8 (\textit{right; solid line}),
$S_\mathrm{a}$ at freezeout, $S_\mathrm{a,f}$, in each $R_\mathrm{s}$
case is shown as a function of $Y_e$. The $r$-process paths predicted
from the $(n, \gamma)$-$(\gamma, n)$ approximation \citep{Gori96},
\begin{equation}
S_a^0 ({\rm MeV})
\equiv \left(34.075 - \log N_n + \frac{3}{2} \log T_9 \right)
\frac{T_9}{5.04},
\end{equation}
are also shown in Figure~8 (\textit{right; dotted line}) for
comparison purposes.

As seen in Figure~8 (\textit{top right}), the $(n, \gamma)$-$(\gamma,
n)$ approximation can reasonably predict the $r$-process paths at the
freezeout for $R_\mathrm{s} \le 1000$~km, in which $S_\mathrm{a}^0$ is
systematically ($0.5-1.0$~MeV) higher than $S_\mathrm{a,f}$. This is a
consequence that the classical $r$-process conditions
$\tau_{n\gamma}/\tau_{\gamma n} \approx 1$ and $\tau_\beta/\tau_n \gg
1$ are approximately valid in these cases owing to the sufficiently
high temperatures and densities. In contrast, the $(n,
\gamma)$-$(\gamma, n)$ approximation poorly predicts the \textit{real}
path at the freezeout for $R_\mathrm{s} \ge 3000$~km. This is due to
the low temperatures $T_\mathrm{9,f} < 0.7$ as well as the low
densities (Fig.~5) in these cases, in which the $(n,
\gamma)$-$(\gamma, n)$ equilibrium is not valid \citep[``cold
$r$-process'',][]{Wana07}. The classical conditions above are replaced
by $\tau_{n\gamma}/\tau_{\gamma n} \ll 1$ and $\tau_\beta/\tau_n
\gtrsim 1$. The latter condition, that is, the competing $\beta$-decay
with the neutron capture, pushes the $r$-process path $S_\mathrm{a}$
back to a higher value \citep[see the similar situation in the
decomposition of cold neutron star matter,][]{Latt77, Meye89, Frei99,
Gori05b}. As a consequence, the $R_\mathrm{s} \ge 3000$~km cases take
similar $S_\mathrm{a,f}$ values. This is the reason that the
$R_\mathrm{s} \le 1000$~km cases result in quite different abundance
curves, while the other cases do not show significant differences. In
addition, $S_\mathrm{a,f}$ values for $R_\mathrm{s} = 1000$~km and
$\ge 10000$~km are close at $Y_e \sim 0.25$. As a result, the
abundance curves around the third peak in these cases are similar
(Fig.~7). Figure~8 (\textit{top panels}) shows that the requisite
conditions for the second and third peak abundances are
$T_\mathrm{9,e} = 1.2-1.4$ and $\le 1.0$, respectively, which result
in $S_\mathrm{a,f} = 4.2-4.4$~MeV and $2.8-3.8$~MeV.

In summary, in the early winds ($L_{\nu, 51} = 10$), the termination
shock can play a decisive role for the second and rare-earth peak
abundances. Only a rather small $R_\mathrm{s}$ ($\sim 500$~km) is
allowed to reproduce the solar $r$-process curve in this atomic mass
range ($A \approx 100-180$). The termination shock may have a role for
the third peak abundances as well (e.g., $R_\mathrm{s} =
1000$~km). The solar $r$-process curve in this mass range ($A \approx
180-200$) is, however, well reproduced without shock (e.g.,
$R_\mathrm{s} = 30,000$~km), too. Any subsonic winds cannot reproduce
the solar $r$-process curve at all, in which $n$-exhaustion occurs at
rather high temperature (Fig.~5; \textit{top right}).

In the late winds ($L_{\nu, 51} = 1$), all the nucleosynthetic results
for $R_\mathrm{s} \ge R_\mathrm{c} = 1056$~km cannot be distinguished
(not displayed in this paper), which have similar abundance curves to
those in the early winds ($L_{\nu, 51} = 10$) with $R_\mathrm{s} \ge
10,000$~km (Figs.~6 and 7). This is due to the quite low temperature
($T_9 = 0.27$, Fig.~1) at the sonic point. As a consequence, freezeout
has already occurred at $r < R_\mathrm{s}$ in most cases (Fig.~8;
\textit{bottom left}), in which the termination-shock has no effect on
the $r$-process. This leads to almost the same $S_\mathrm{a,f}$ curves
(Fig.~8; \textit{bottom right}), which are similar to those in the
early wind ($L_{\nu, 51} = 10$) with $R_\mathrm{s} \ge 10,000$~km
(Fig.~8; \textit{top right}). The resulting $S_\mathrm{a,f}$ values
($2.9-3.9$~MeV) fall within the requisite range to reproduce the solar
$r$-process curve near the third peak ($A \approx 180-200$), but too
small for the second and rare-earth peaks. Therefore, the late winds
can be responsible only for the third peak abundances and heavier, in
which the effect of termination shock is of no importance. It should
be noted that subsonic winds may be able to reproduce the second and
rare-earth peak abundances, if $T_\mathrm{9,e}$ appears to be
$1.2-1.4$ (Fig.~1; \textit{top right}).

\subsection{Effect of Termination-Shock Heating}

As noted in \S~3, the large entropy jumps owing to the
termination-shock heating do not help to increase $Y_n/Y_h$, since the
jumps appear only when the temperature decreases below $T_9 = 2.5$
(Fig.~3). In fact, all the $R_\mathrm{s}$ cases for a given
($M_\mathrm{NS}$, $L_\nu$) set have the identical $Y_n/Y_h$ curve as a
function of $Y_e$ in Figure~4. One may consider, however, a large
entropy jump during $r$-processing affects the nucleosynthetic path
and modifies the final abundance curve. An entropy increase is
equivalent to a reduction of the matter density (and thus the neutron
number density) at a fixed temperature in the current radiation
dominated outflows. A sizable entropy jump can thus, in principle,
modify the $r$-process path.

Surprisingly, even the entropy jump to about $400\, N_\mathrm{A} k$
does not change the final abundance curves. In Figures~6 and 7, the
dotted lines denote the nucleosynthetic results by suppressing the
entropy jumps without changing the temperature histories (but
artificially increase the densities). No differences between the
abundance curves with and without entropy jumps can be seen in
Figure~6, and only minor differences appear for $R_\mathrm{s} =
3000$~km and 10,000~km in Figure~7.

The reason is that the entropy jump is too small in the $R_\mathrm{s}
\le 1000$~km cases (Fig.~3) to change the $r$-process path during the
freezeout phase. On the other hand, a large entropy jump appears only
at late times when the $r$-process has already ceased in most
cases. As an example, for $R_\mathrm{s} = 30,000$~km, all the $Y_e \ge
0.20$ cases experience $n$-exhaustion before the termination-shock
heating (Fig.~8, \textit{top left}). For $R_\mathrm{s} = 3000$~km and
10,000~km, the winds with $Y_e \le 0.29$ and $Y_e \le 0.21$,
respectively, reach the termination shock radii before
$n$-exhaustion. These cases, however, result in only minor
modifications of the final abundance curves (Fig.~7).

\section{Conclusions and Discussion}

We have investigated the role of wind termination shock on the
$r$-process nucleosynthesis in supersonically expanding
neutrino-driven outflows. For this purpose, a semi-analytic, general
relativistic neutrino-driven wind model was utilized to obtain
spherically symmetric transonic wind solutions. The Rankine-Hugoniot
relations were applied at arbitrary chosen shock radii $R_{s}$ to
obtain the boundary conditions for the subsequent termination-shocked
subsonic outflows.

We explored the properties of termination-shocked supersonic winds for
various neutron star masses $M_\mathrm{NS}$, neutrino luminosities
$L_\nu$, and termination-shock radii $R_\mathrm{s}$. $M_\mathrm{NS}$
was taken to be $1.4\, M_\odot$ and $2.0\, M_\odot$ as a typical and a
most massive proto-neutron stars, respectively. $L_{\nu, 51}$ was
taken to be 10 and 1 as representative of early and late winds,
respectively. The earlier winds with $L_{\nu, 51} > 10$, in which
matter is likely to be proton-rich \citep{Bura06, Froh06, Kita06}, may
not be relevant for the $r$-process. The later winds with $L_{\nu, 51}
< 1$ may not contribute to the Galactic production of $r$-process
nuclei owing to their small mass ejection rates \citep[$\la 10^{-6}\,
M_\odot$,][]{Wana01, Thom01}.

\subsection{Conclusions}

$R_\mathrm{s}$ was changed from the sonic radius $R_\mathrm{c}$ ($\sim
200-1000$~km) up to 30,000~km. Our main results are summarized as
follows.

1. The shock jumps are greater at larger $R_\mathrm{s}$ from the sonic
point, for fixed $M_\mathrm{NS}$ and $L_\nu$. The termination-shocked
winds become, by definition, subsonic outflows, in which the
temperature and density decrease rather slowly with radius (and with
time owing to the decaying expansion velocity). The entropy increases
by the termination-shock heating up to several $100\, N_\mathrm{A} k$
in some extreme cases ($R_\mathrm{s} = 30,000$~km), which is several
times larger than its unshocked value. The jump is, on the other hand,
only moderate at small $R_\mathrm{s}$ ($\sim 1000$~km), which
disappears in the (subsonic) $R_\mathrm{s} = R_\mathrm{c}$ case.

2. For fixed $M_\mathrm{NS}$ and $R_\mathrm{s}$, the shock jump is
   more prominent in the earlier wind ($L_{\nu, 51} = 10$) owing to
   its faster wind velocity. In the early wind, the shock jump occurrs
   even at $R_\mathrm{s} =$ several 100~km when the temperature is as
   high as $T_9 \ga 1$, because of its smaller sonic radius
   ($R_\mathrm{c} \sim 200$~km). On the other hand, the jump occurrs
   only at low temperature ($T_9 \la 0.2$) in the late wind ($L_{\nu,
   51} = 1$) owing to its distant sonic radius ($R_\mathrm{c} \sim
   1000$~km). For fixed $L_\nu$ and $R_\mathrm{s}$ (as well as for a
   fixed $R_\mathrm{NS} = 10$~km), the jump is significantly greater
   in the massive $M_\mathrm{NS}$ ($= 2.0\, M_\odot$) case owing to
   its somewhat faster wind velocity.

3. The entropy jump takes place only after the temperature decreases
   below $T_9 \approx 2.5$, except for extremely high
   $L_\nu$. Therefore, the termination-shock has no effect on
   increasing the neutron-to-seed ratio.

All these results are consistent, at least qualitatively, with those
found in one-dimensional hydrodynamic simulations by
\citet{Arco07}. We further performed detailed nucleosynthesis
calculations to examine in particular the effects of shock
deceleration and shock heating on the $r$-process.

4. The temperature in the shock decelerated wind decreases rather
slowly with time. As a result, the temperature stays mostly constant
during the $r$-process freezeout phase \citep[as assumed in][]{Wana02,
  Wana07}. We find that the decelerated temperature is highly
dependent on $L_\nu$ and $R_\mathrm{s}$. In the early wind ($L_{\nu,
  51} = 10$), the termination-shock deceleration plays a decisive role
to determine the final $r$-process curve. A solar-like $r$-process
curve including the second and rare-earth peaks ($A \approx 100-180$)
can be obtained with rather small $R_\mathrm{s}$ ($\sim 500$~km), in
which the $r$-process freezeout takes place at $T_9 \approx
1.2-1.4$. On the other hand, the heaviest portion of the solar
$r$-process curve ($A \approx 180-200$) can be reproduced both with
$R_\mathrm{s} \sim 1000$~km ($T_9 \approx 1.0$) and $R_\mathrm{s} \ga
10,000$~km ($T_9 \la 0.3$). In the latter cases, the $r$-process has
ceased before the shock deceleration, in which the termination shock
plays no role. In the late winds ($L_{\nu, 51} = 1$), the shock
deceleration can occur only at $R_\mathrm{s} \ge 1000$~km, in which
the $r$-process has already frozen out in most cases. In the late
winds, therefore, only the heaviest part of the solar $r$-process
curve can be reproduced, in which the termination shock has little
effect.

5. The entropy jumps by the termination-shock heating have negligible
   effects on the final $r$-process curve. This is a consequence that
   a sizable entropy jump occurs only for $R_\mathrm{s} \ga 10000$~km
   when the temperature is as low as $T_9 \la 0.3$, where the
   $r$-process has already ceased. For smaller $R_\mathrm{s}$, the
   shock heating can occur before freezeout, but the entropy jump is
   too small to modify the $r$-process path.

\subsection{Discussion}

Our approach was based on the simple steady state wind solutions with
arbitrary chosen (but reasonable) input parameters, although our
results are qualitatively consistent with more detailed hydrodynamic
results by \citet{Arco07}. We also employed rather low initial $Y_e$
($=0.2-0.4$) to obtain the $r$-process abundances with reasonable
parameter settings. Our results presented here should be therefore
taken to be only suggestive. Keeping this caution in mind, we suggest
a couple of possible supernova progenitors that may be relevant for
the astrophysical $r$-process site.

First is a rather massive progenitor ($\sim 25\, M_\odot$), in which
the termination-shock radius may stay close to the sonic point for
long time. \citet{Arco07} show that a more massive progenitor has a
higher mass accretion rate that results in a higher $L_\nu$ and a
smaller $R_\mathrm{s}$. As a consequence, in their $25\, M_\odot$
case, the temperature stayed at $T_9 \ga 1.0$ during the first 10
seconds after core bounce. In this condition, both the lighter ($A
\approx 100-180$) and heavier ($A \approx 180-200$) parts of the solar
$r$-process curve would be reproduced if requisite neutron-to-seed
ratios were achieved. This condition $T_9 \ga 1.0$ (as well as high
neutron number density, $> 10^{20}$~cm$^{-3}$) has in fact long been
considered as a physical requirement to account for the solar
$r$-process abundance curve \citep[e.g.,][]{Math90, Krat93}. A massive
progenitor forms a massive proto-neutron star, in which general
relativistic effects may in part help to increase entropy and reduce
expansion timescale \citep{Card97, Otsu00, Wana01, Thom01}.

Second is, in contrast, a star of $\sim 10\, M_\odot$ near the
low-mass end of supernova-progenitor range. \citet{Arco07} show that,
in their $10.2\, M_\odot$ case, the termination shock propagated
outward quickly owing to the steep density gradient, in which the
temperature at $R_\mathrm{s}$ dropped to $T_9 \sim 0.1$ after 10
seconds. In this case, the lighter part ($A \approx 100-180$) of the
solar $r$-process curve should be formed in shock-decelerated early
winds, in which the (shocked) temperature is still as high as $T_9
\approx 1.2-1.4$. The heavier part may be followed in later winds, in
which the $r$-process proceeds in very low temperatures \citep[cold
$r$-process,][]{Wana07}. Low-mass supernovae have been also suggested
as the astrophysical $r$-process site, from Galactic chemical
evolution studies \citep[$\sim 8-10\, M_\odot$,][]{Math90, Ishi99,
Ishi04}. The mechanism to obtain requisite physical conditions for the
$r$-process in such a supernova, however, has been lacking.

Needless to say, we cannot exactly constrain the mass range of
supernova progenitors on the basis of our current (only suggestive)
results. It is likely, however, that the (shocked) temperatures in
intermediate progenitor cases (e.g., $15-20\, M_\odot$) stay at $T_9
\sim 0.5-0.7$ for long time, in which the third peak abundances may be
lower-shifted compared to the solar $r$-process curve \citep[Fig.~7,
see also][]{Wana02, Wana07}. It should be also noted that the subsonic
outflow has higher temperature than that in the transonic wind at any
radii. We cannot exclude, therefore, possibilities that these
progenitors provide suitable temperature $T_9 \approx 1.0-1.4$ during
the $r$-process. \citet{Arco07} show, however, that all their
simulations with ``standard'' parameter settings resulted in forming
supersonic outflows within 10 seconds after core bounce. Only a subset
of simulations with somewhat ``extreme'' parameter choices appeared to
have the transiting features between transonic and subsonic outflows.

Given that our speculations above are correct, it is still difficult
to specify which of $\sim 10\, M_\odot$ or $\sim 25\, M_\odot$ stars
are the dominant sources of the Galactic $r$-process nuclei. Both
cases can potentially reproduce the whole mass range of the solar
$r$-process curve \citep[but Pb abundance can be substantially
different, see Fig.~7 and][]{Wana07}. Detailed nucleosynthesis
calculations in time-evolving neutrino-driven winds with the
termination shock will be needed to draw more definitive answers. It
should be also noted that we assumed only spherically symmetric
outflows in this study, while recent hydrodynamic studies suggest some
decisive roles of multidimensional effects on the mechanism of
neutrino-driven explosions \citep[e.g.,][]{Bura06, Ohni06,
Burr07}. Eventually, we will need extensive nucleosynthesis studies in
the outflows obtained from multidimensional core-collapse simulations.

\acknowledgements We would like to thank an anonymous referee for
valuable comments.  This work was supported in part by a Grant-in-Aid
for Scientific Research (17740108, 18104003, 18540231), and from the
21st Century COE Program (QUEST) from the Ministry of Education,
Culture, Sports, Science, and Technology of Japan.

\clearpage

\begin{figure}
\epsscale{1.}
  \plotone{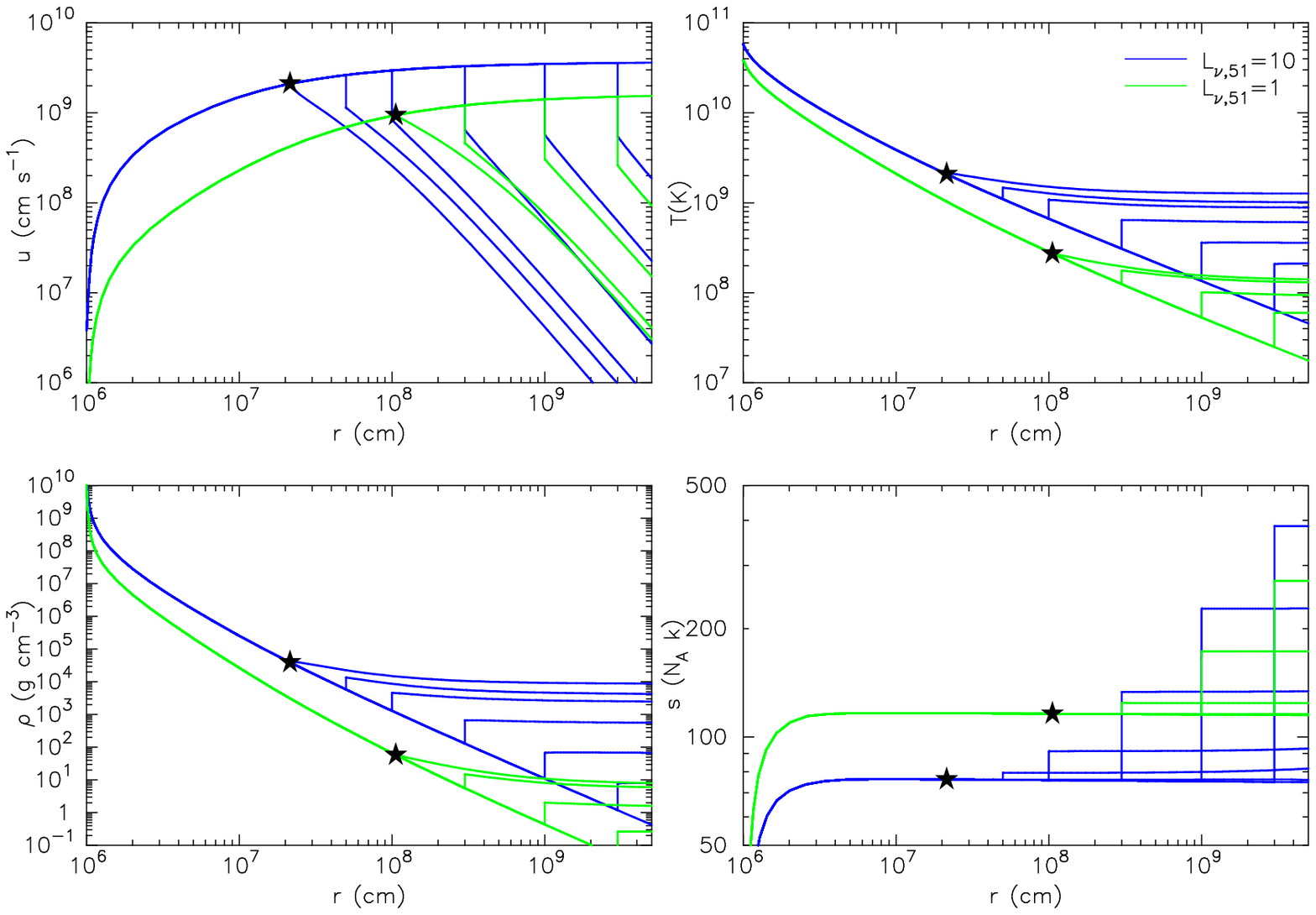}
  \caption{Radial velocity $u$ (\textit{top left}), temperature $T$
(\textit{top right}), density $\rho$ (\textit{bottom left}), and
entropy $s$ (\textit{bottom right}) in neutrino-driven winds for
$M_\mathrm{NS} = 1.4\, M_\odot$ as functions of radius. The neutrino
luminosities are taken to be $L_{\nu,51}=10$ (\textit{blue}) and 1
(\textit{green}), respectively. In each $L_\nu$ case, the
termination-shock radius is varied from the sonic radius (denoted by
\textit{star}) to infinity (see text).}
\end{figure}

\clearpage

\begin{figure}
\epsscale{1.}
  \plotone{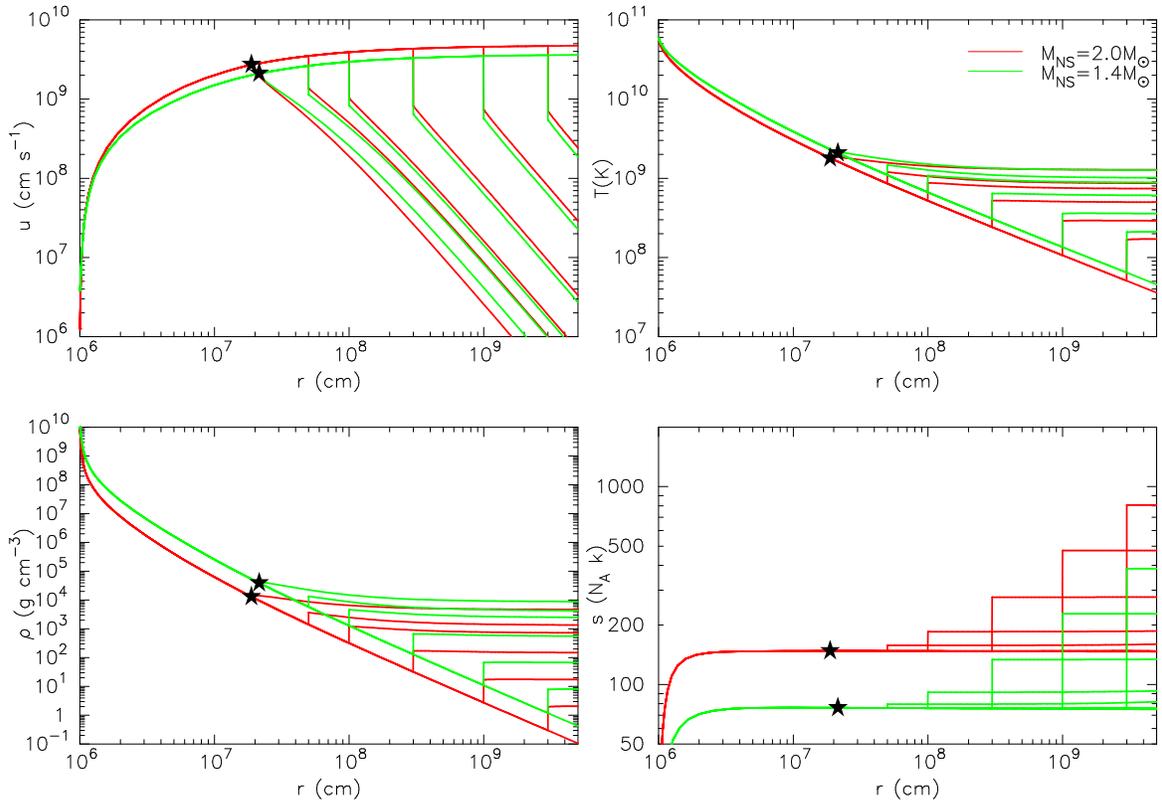}
  \caption{Same as Figure~1, but for $M_\mathrm{NS} = 1.4\, M_\odot$
(\textit{green}) and $2.0\, M_\odot$ (\textit{red}). The neutrino
luminosity is taken to be $L_{\nu,51}=10$.}
\end{figure}

\clearpage

\begin{figure}
\epsscale{1.}
  \plotone{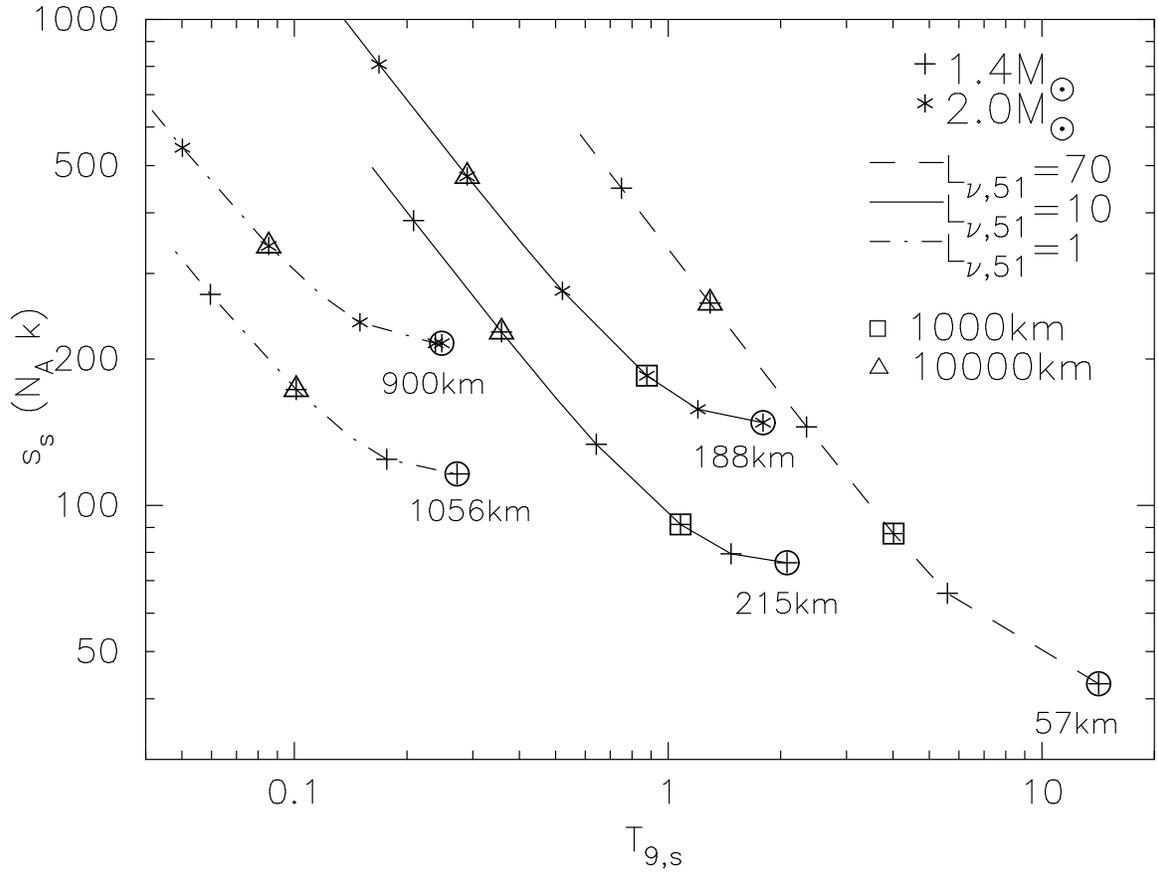}
  \caption{Relations between the entropies $s_{\rm s}$ and the
  temperatures $T_{\rm s}$ just above the termination-shock radii
  $R_\mathrm{s}$. The neutron star masses and the neutrino
  luminosities are taken to be $M_{\rm NS}=1.4\, M_\odot$
  (\textit{crosses}), $2.0\, M_\odot$ (\textit{asterisks}) and
  $L_{\nu,51}=10$ (\textit{solid lines}), 1 (\textit{dash-dotted
  lines}), respectively. For $M_\mathrm{NS} = 1.4\, M_\odot$, $L_{\nu,
  51} = 70$ (\textit{dashed line}) is also considered as an extreme
  case. $R_\mathrm{s}$ in each $L_\nu$ case is varied from
  $R_\mathrm{c}$ (sonic radius, marked by \textit{circles}) to 500~km,
  1000~km (marked by \textit{squares}), 3000~km, 10000~km (marked by
  \textit{triangles}), and 30000~km, whenever $R_\mathrm{s} >
  R_\mathrm{c}$ is satisfied.}
\end{figure}

\clearpage

\begin{figure}
\epsscale{1.}
\plotone{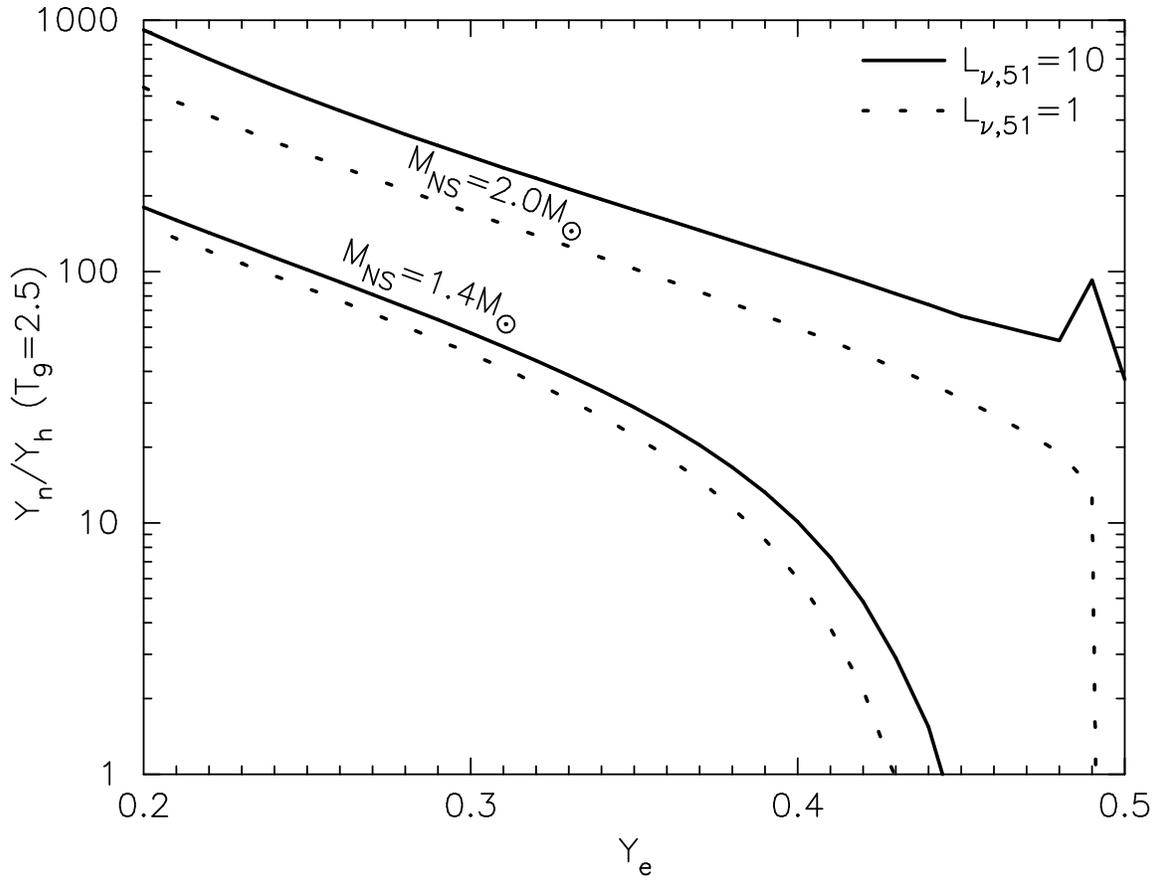}

\caption{Neutron-to-seed ratios $Y_n/Y_h$ at the beginning of
$r$-processing ($T_9 = 2.5$) in $M_\mathrm{NS} = 1.4\, M_\odot$ and
$2.0\, M_\odot$ cases as functions of the initial electron fraction
$Y_e$. Solid and dotted lines denote the results for $L_{\nu, 51}=10$
and 1, respectively.}

\end{figure}

\clearpage

\begin{figure}
\epsscale{1.}
\plotone{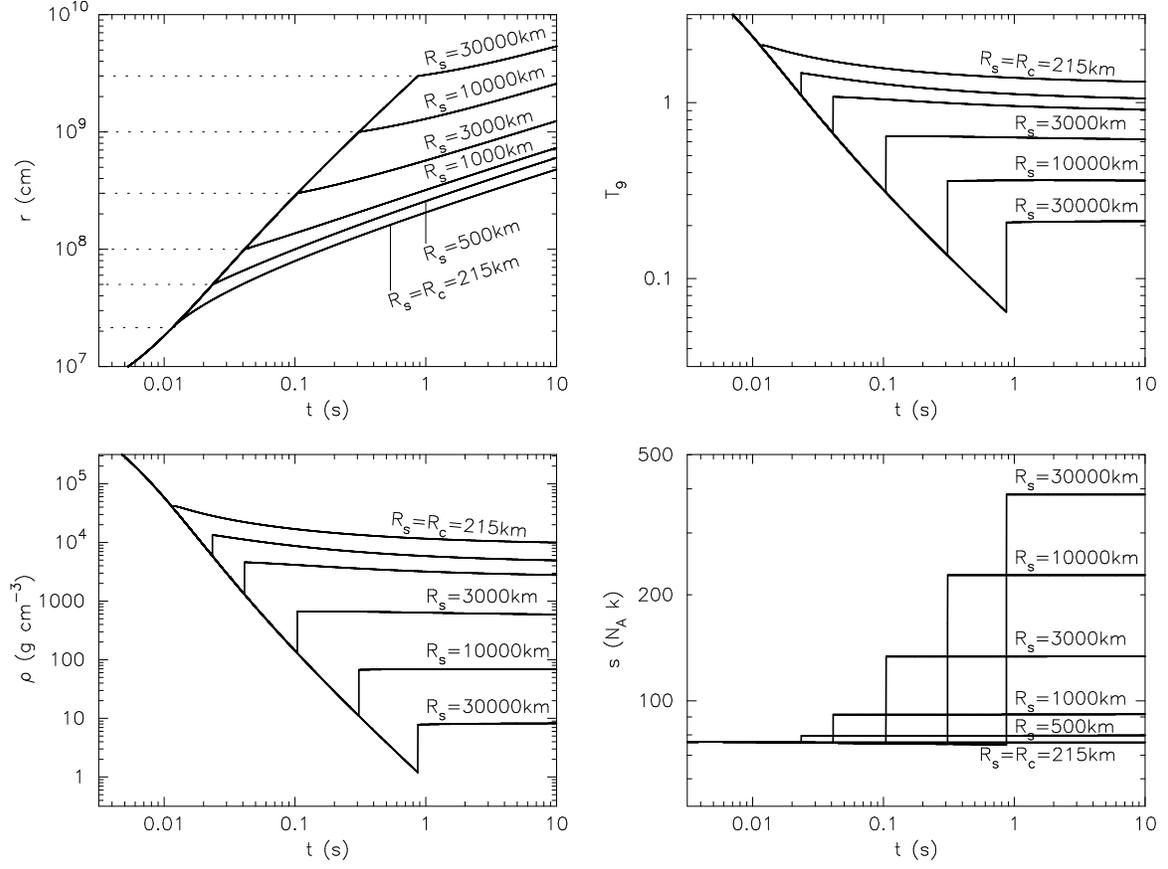}

\caption{Time variations ($t = 0$ at $T_9 = 9$) of $r$ (\textit{top
left}), $T_9$ (\textit{top right}), $\rho$ (\textit{bottom left}), and
$s$ (\textit{bottom right}) for the selected shock radii
$R_\mathrm{s}$. The neutron star mass and the neutrino luminosity are
taken to be $M_\mathrm{NS} = 1.4\, M_\odot$ and $L_{\nu, 51}=10$,
respectively.}

\end{figure}

\clearpage

\begin{figure}
\epsscale{1.}
\plotone{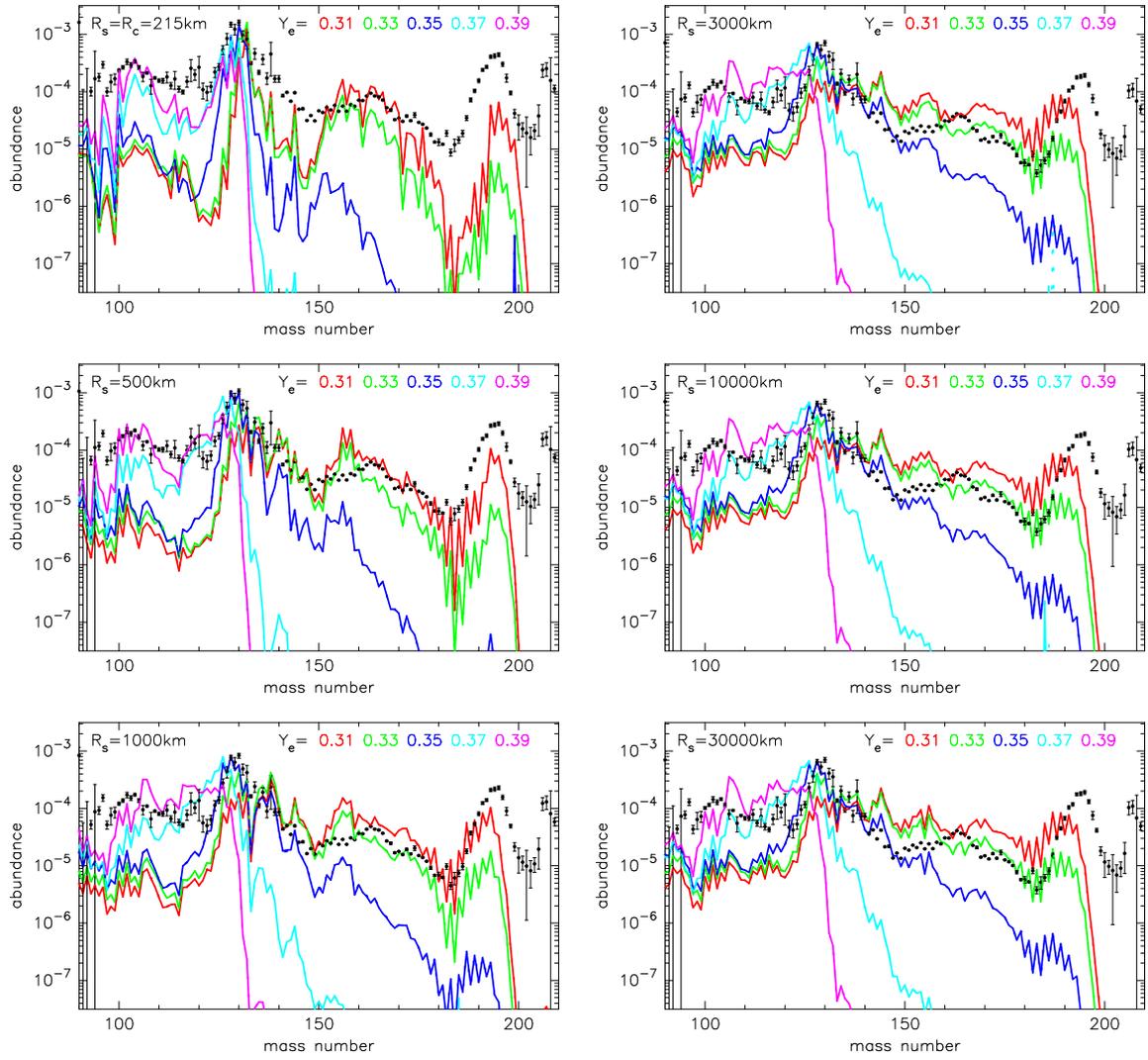}

\caption{Comparison of the nucleosynthetic yields (\textit{lines}) for
  $Y_e = 0.31$, 0.33, 0.35, 0.37, and 0.39 in the $(M_\mathrm{NS},
  L_{\nu, 51}) = (1.4\, M_\odot, 10)$ case with the solar $r$-process
  abundances \citep[\textit{dots};][]{Kapp89} as functions of mass
  number. The latter is shifted to match the height of the second peak
  ($A \approx 130$). }

\end{figure}

\clearpage

\begin{figure}
\epsscale{1.}
\plotone{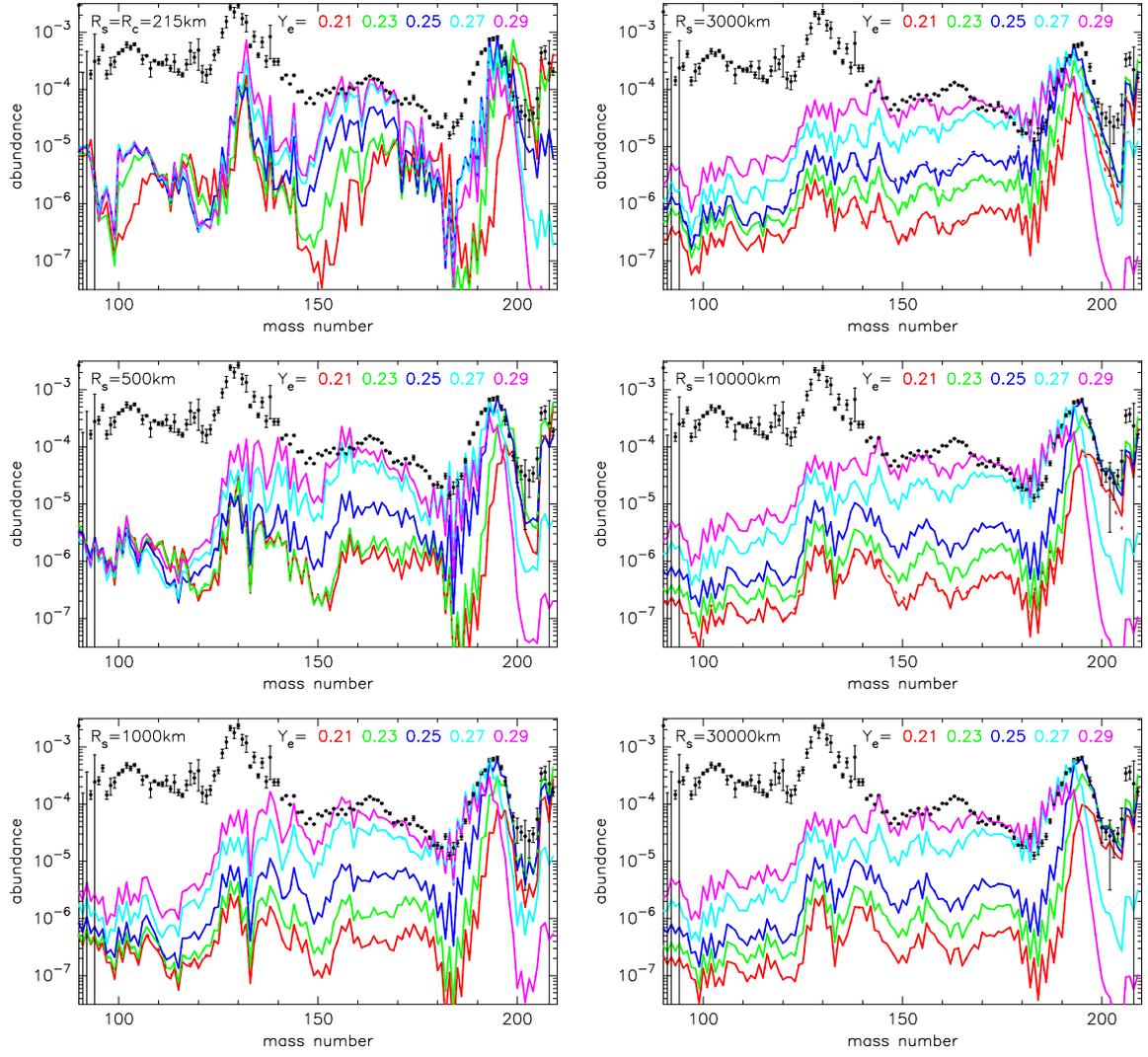}

\caption{Same as Figure~5, but for $Y_e = 0.21$, 0.23, 0.25, 0.27, and
  0.29. The solar $r$-process abundances are shifted to match the
  height of the third peak ($A \approx 195$).}

\end{figure}

\clearpage

\begin{figure}
\epsscale{1.}
\plotone{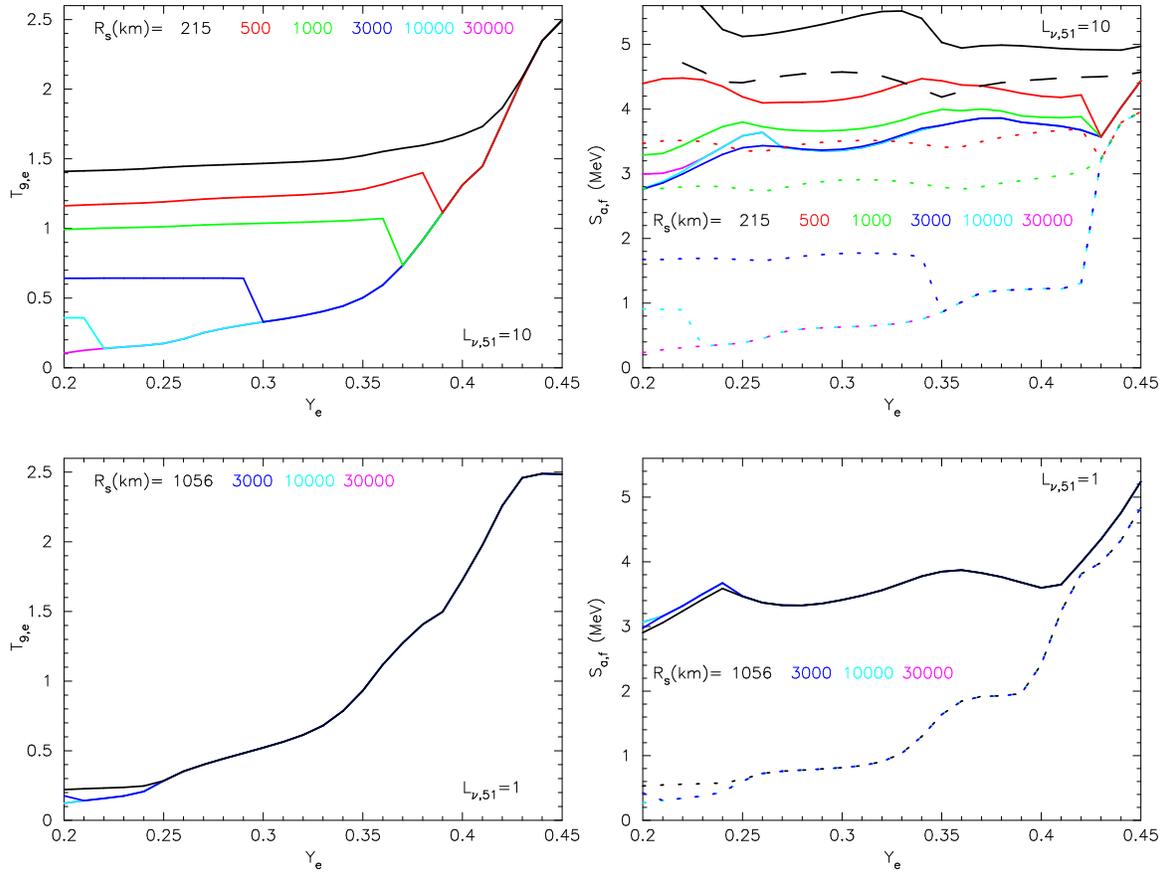}

\caption{Temperatures at $n$-exhaustion $T_\mathrm{9,e}$
(\textit{left}) and nucleosynthetic paths at freezeout (in terms of
neutron separation energies) $S_\mathrm{a,f}$ (\textit{right}) for
$M_\mathrm{NS} = 1.4\, M_\odot$ as functions of $Y_e$. The
termination-shock radius $R_\mathrm{s}$ is varied from the sonic
radius up to 30000~km. Top and bottom panels show the results for
$L_{\nu, 51}=10$ and 1, respectively. The paths predicted from the
$(n, \gamma)$-$(\gamma, n)$ equilibrium are also shown by dotted lines
(\textit{right panels}).}

\end{figure}

\end{document}